\documentclass[acmsmall]{acmart}

\usepackage{hyperref}

\usepackage{wasysym}
\usepackage{textcomp}
\usepackage{xcolor}
\usepackage{url}
\usepackage{utfsym}
\usepackage{enumitem}
\usepackage{pifont}
\usepackage{subcaption}
\usepackage{graphicx}
\usepackage{ulem}
\usepackage{multirow}
\usepackage{listings}
\usepackage{color}
\definecolor{pblue}{rgb}{0.13,0.13,1}
\definecolor{pgreen}{rgb}{0,0.5,0}
\definecolor{pred}{rgb}{0.9,0,0}
\definecolor{pgrey}{rgb}{0.46,0.45,0.48}
\usepackage{pifont}
\usepackage{url}
\usepackage{caption}
\usepackage{bigstrut,bigdelim}
\usepackage{float}  

\usepackage{multirow,multicol}
\usepackage[most]{tcolorbox}
\usepackage{enumitem}
\usepackage{colortbl}
\definecolor{mygray}{gray}{.9}
\usepackage{float}

\usepackage{xspace}
\usepackage{tikz}
\usepackage{pgfplots}
\usepackage{colortbl}
\pgfplotsset{compat=1.3}

\makeatletter
\DeclareRobustCommand\onedot{\futurelet\@let@token\@onedot}
\def\@onedot{\ifx\@let@token.\else.\null\fi\xspace}

\def\method{{\sc ProCoder}\xspace}

\usepackage{multirow}
\usepackage[most]{tcolorbox}
\usepackage{pifont}
\usepackage{graphicx}

\usepackage{xspace}
\usepackage{enumitem}
\usepackage{colortbl}
\usepackage{pifont}
\usepackage{circledsteps}

\usepackage{xcolor}
\usepackage{pifont}
\usepackage{enumitem}
\usepackage{algorithm}
\usepackage{algpseudocode}
\usepackage{graphicx}
\usepackage{array} 
\usepackage{multirow}
\usepackage[table]{xcolor}
\usepackage{colortbl}
\usepackage{threeparttable}
\usepackage[most]{tcolorbox}
\usepackage{listings}
\usepackage{lipsum}
\usepackage{breqn} 
\tcbuselibrary{listingsutf8}
\usepackage{nameref}
\tcbuselibrary{listings,breakable}

\usepackage[framemethod=tikz]{mdframed}
\mdfdefinestyle{customstyle}{
  linecolor=gray!100,
  linewidth=3pt,
  innerleftmargin=3pt,
  topline=false,
  rightline=false,
  bottomline=false,
  leftline=true,
  innerrightmargin=3pt,
  innertopmargin=3pt,
  innerbottommargin=3pt,
  backgroundcolor=gray!15
}
\newenvironment{custommdframed}
  {\begin{mdframed}[style=customstyle]}
  {\end{mdframed}}

\newcommand{\cmark}{\textcolor{green!60!black}{\ding{51}}} 
\newcommand{\xmark}{\textcolor{red}{\ding{55}}} 
\definecolor{DeepGreen}{RGB}{0,128,0}
\definecolor{DeepRed}{RGB}{178,34,34}
\definecolor{lightgray}{gray}{0.92}

\definecolor{codegreen}{rgb}{0,0.6,0}
\definecolor{codegray}{rgb}{0.5,0.5,0.5}
\definecolor{codepurple}{rgb}{0.58,0,0.82}
\definecolor{backcolour}{rgb}{0.95,0.95,0.96}
\definecolor{framecolor}{rgb}{0.8,0.8,0.8}

\def\textbfmethod{{\sc \textbf{GraphCodeAgent}}\xspace}
\def\method{{\sc GraphCodeAgent}\xspace}

\newcommand{\eg}{{\emph{e.g.,}}\xspace}
\newcommand{\ie}{{\emph{i.e.,}}\xspace}
\newcommand{\etc}{{\emph{etc}}\xspace}

\newcommand{\increase}[1]{\textcolor{red}{ #1}}

\AtBeginDocument{%
  \providecommand\BibTeX{{%
    \normalfont B\kern-0.5em{\scshape i\kern-0.25em b}\kern-0.8em\TeX}}}

\setcopyright{acmcopyright}
\copyrightyear{2018}
\acmYear{2018}
\acmDOI{XXXXXXX.XXXXXXX}

\acmJournal{JACM}
\acmVolume{37}
\acmNumber{4}
\acmArticle{111}
\acmMonth{8}

\begin{document}

\title{\textbfmethod: Dual Graph-Guided LLM Agent for Retrieval-Augmented Repo-Level Code Generation}

\author{Jia Li}
\email{jia.li@whu.edu.cn}
\affiliation{
  \institution{School of Computer Science, Wuhan University}
  \city{Wuhan}
  \country{China}
}

\author{Xianjie Shi}
\email{2100013180@stu.pku.edu.cn}
\author{Kechi Zhang}
\email{zhangkechi@pku.edu.cn}
\affiliation{
  \institution{Key Lab of High Confidence Software Technology (PKU), Ministry of Education School of Computer Science, Peking University}
  \city{Beijing}
  \country{China}
}

\author{Ge Li}
\authornote{Corresponding author}
\email{lige@pku.edu.cn}
\affiliation{%
  \institution{
  Key Lab of High Confidence Software Technology (PKU), Ministry of Education School of Computer Science, Peking University
  }
  \city{Beijing}
  \country{China}
  }

\author{Zhi Jin}
\authornotemark[1]
\email{zhijin@whu.edu.cn}
\affiliation{
  \institution{School of Computer Science, Wuhan University}
  \city{Wuhan}
  \country{China}
}

\author{Lei Li}
\affiliation{%
  \institution{
  The University of Hong Kong
  }
  \city{Hong Kong}
  \country{China}
  }
\email{nlp.lilei@gmail.com}

\author{Huangzhao Zhang}
\affiliation{%
  \institution{
  Independent
  }
  \city{Beijing}
  \country{China}
  }
\email{zhang_hz@pku.edu.cn}

\author{Jia Li}
\affiliation{%
  \institution{
   College of AI, Tsinghua University
  }
  \city{Beijing}
  \country{China}
  }
\email{jia\_li@mail.tsinghua.edu.cn}

\author{Fang Liu}
\affiliation{%
  \institution{
  Beihang University
  }
  \city{Beijing}
  \country{China}
  }
\email{fangliu@buaa.edu.cn}

\author{Yuwei Zhang}
\affiliation{%
  \institution{
  Institute of Software Chinese Academy of Sciences
  }
  \city{Beijing}
  \country{China}
  }
\email{zhangyuwei@iscas.ac.cn}

\author{Zhengwei Tao}
\affiliation{%
  \institution{
  Peking University
  }
  \city{Beijing}
  \country{China}
  }
\email{tttzw@stu.pku.edu.cn}

\author{Yihong Dong}
\email{dongyh@stu.pku.edu.cn}
\affiliation{
  \institution{Key Lab of High Confidence Software Technology (PKU), Ministry of Education School of Computer Science, Peking University}
  \city{Beijing}
  \country{China}
}

\author{Yuqi Zhu}
\affiliation{%
  \institution{
  Academy of Military Sciences
  }
  \city{Beijing}
  \country{China}
  }
\email{zhuyuqi1997@126.com}

\author{Chongyang Tao}
\affiliation{%
  \institution{
  Beihang University
  }
  \city{Beijing}
  \country{China}
  }
\email{chongyang@buaa.edu.cn}


\begin{abstract}

Writing code requires significant time and effort in software development. To automate this process, researchers have made substantial progress for code generation. 
Recently, large language models (LLMs) have demonstrated remarkable proficiency in function-level code generation, yet their performance significantly degrades in the real-world software development process, where coding tasks are deeply embedded within specific repository contexts. 
Existing studies attempt to use retrieval-augmented code generation (RACG) approaches to mitigate this demand by providing LLMs with retrieved code snippets. However, there is a gap between natural language requirements and programming implementations, where a natural language requirement typically states a high-level goal, yet implicitly consists of multiple fine-grained functional elements that are not directly expressed.
This results in the failure to retrieve the relevant code of these fine-grained subtasks through existing retrieval methods, which largely rely on lexical or semantic matching between requirements and code. 
To address this challenge, we propose \method, a dual graph-guided LLM agent for retrieval-augmented repo-level code generation, bridging the gap between natural language requirements and programming implementations. Our approach constructs two interconnected graphs: a Requirement Graph (RG) to model requirement relations of code snippets within the repository, as well as the relations between the target requirement and the requirements of these code snippets, and a Structural-Semantic Code Graph (SSCG) to capture the repository's intricate code dependencies. 
Guided by this, an LLM-powered agent performs multi-hop reasoning to systematically retrieve all context code snippets, including implicit code such as invoked predefined APIs and multi-hop related code snippets, and explicit code snippets such as semantically similar code and up-to-date domain knowledge through web searching, even if they are not explicitly expressed in requirements.
We evaluated \method on three advanced LLMs with the two widely-used repo-level code generation benchmarks DevEval and CoderEval. Extensive experiment results show that \method significantly outperforms state-of-the-art baselines, achieving relative improvements of 43.81\% with GPT-4o and 39.15\% with Gemini-1.5-Pro on DevEval, and 31.91\% with GPT-4o and 8.25\% with Gemini-1.5-Pro on CoderEval in terms of Pass@1. 
The advantage of \method is particularly significant when handling non-standalone code with complex dependencies, highlighting its potential for practical application in complex software development workflows.

\end{abstract}



\received{20 February 2007}
\received[revised]{12 March 2009}
\received[accepted]{5 June 2009}

\maketitle

\section{introduction}

\begin{figure*}[t] 
\centering
\includegraphics[width=0.8\linewidth]{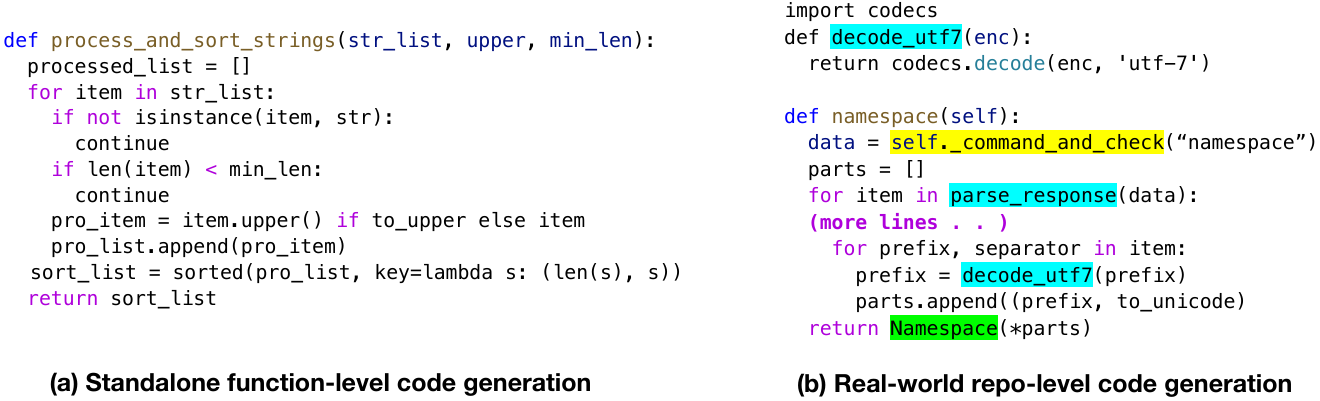}
\caption{Comparison between standalone function-level code generation and real-world repo-level code generation. Repo-level code generation invokes the code snippets predefined in the repository.}
\label{fig: code generation level}
\end{figure*}

Automating code generation has long captivated the software engineering community \cite{zhang2024codeagent, lozhkov2024starcoder}. 
Recent advancements in large language models (LLMs), such as GPT-4o \cite{GPT-4o}, DeepSeek-V3 \cite{liu2024deepseek}, and Gemini-2.5-Pro \cite{Gemini-2.5-Pro}, have demonstrated impressive capabilities in function-level code generation—achieving Pass@1 rates above 90\% on several benchmarks like HumanEval \cite{chen2021evaluating}. 
However, these results usually do not transfer effectively to real-world software development process, where coding tasks are deeply embedded within specific repository contexts.
In practice, repo-level code generation requires not only syntactic correctness but also awareness of project-specific structure, dependencies, and conventions.
Without such contextual knowledge, the performance of LLMs degrades significantly. 
As illustrated in Figure \ref{fig: code generation level}, different from standalone function-level code generation, repo-level code generation needs to invoke the functions predefined in the repository. This gap highlights the critical need for models to incorporate repository-aware knowledge.

To mitigate this knowledge gap, retrieval-augmented code generation (RACG) \cite{zhang2023repocoder, liu2024graphcoder, li2023acecoder, li2025large, zhang2024codeagent} has become a mainstream strategy, aiming to provide LLMs with relevant context information retrieved from the repository.
Early approaches use sparse retrieval \cite{li2023acecoder} and dense retrieval \cite{li2025large} to search textually similar or semantically similar code from the repository.
Nevertheless, these RACG solutions still have multiple main limitations. 
They treat the repository as multiple standalone code snippets, overlooking structural relationships among them; moreover, dense retrieval relies on maintaining and frequent updating of vectorized representations, resulting in heavy burden for large and evolving repositories.
Lately, researchers \cite{liu2024codexgraph, liu2024graphcoder} incorporate structural information of a repository to refine retrieval through constructing graphs. They typically consider simple structural information, such as basic data flow parsing and dependency analysis, resulting in the failure to fully explore the entire repository. 
Recent agent-based approaches \cite{zhang2024codeagent, chen2025locagent, xia2024agentless} have received increasing attention in code generation, which can iteratively retrieve context code through its powerful tool utilization ability.
But they still struggle to effectively navigate and comprehend complex structures of the repository, particularly when multi-hop reasoning is required to trace from a requirement to its related code regions that are not directly mentioned.

Despite the fact that modern LLMs support context windows of hundreds of thousands of tokens, their ability to understand code decays significantly as input length increases \cite{li2025longcodeu}.
This phenomenon makes LLMs impractical to process an entire complex repository in one pass, underscoring the need for intelligent and selective retrieval of relevant code segments.
Unlike conventional retrieval tasks, which largely rely on lexical or semantic matching between queries and documents \cite{li2019sampling, feng2019learning}, RACG must bridge the gap between natural language requirements and programming implementations.
A natural language requirement typically states a high-level goal, yet implicitly consists of multiple fine-grained functional elements that are not directly expressed. 
For instance, the user requirement of constructing a user registration feature in the library management system requires implementing several subtasks, including user input validation, security checks, and database operations, \etc. The implementations of these subtasks predefined as APIs are usually invoked by the target code of the user requirement. However, these subtasks are not directly mentioned in the requirement, resulting in the failure to retrieve the relevant code of these subtasks through most existing retrieval methods. Moreover, successfully generating the target code requires not only retrieving the implementations of these subtasks above, but also performing multi-step reasoning to identify additional code snippets, such as related classes, variables, or configuration files, which interact with or support the implementations of subtasks.  
This inherent disconnect between natural language requirements and the implicit, distributed structure of code presents a major challenge for most existing RACG approaches \cite{li2023acecoder, li2025large, liu2024codexgraph, liu2024graphcoder, zhang2024codeagent}, which often fail to capture such nuanced, multi-hop dependencies within the repository.

\begin{figure*}[t] 
\centering
\includegraphics[width=0.8\linewidth]{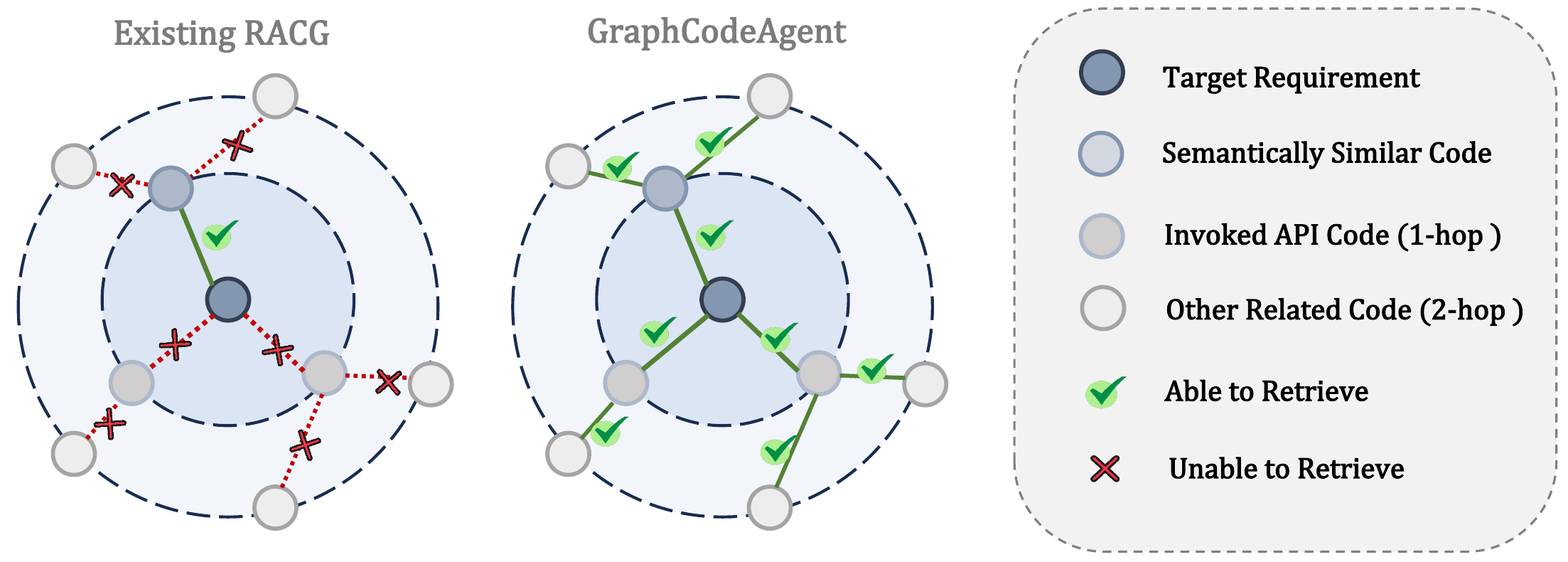}
\caption{ Retrieved knowledge of existing RACG approaches and \method. Our approach can effectively retrieve implicit and explicit knowledge even if they are not directly mentioned in the target requirement.}
\label{fig: reason comparison}
\end{figure*}

\begin{figure*}[t] 
\centering
\includegraphics[width=\linewidth]{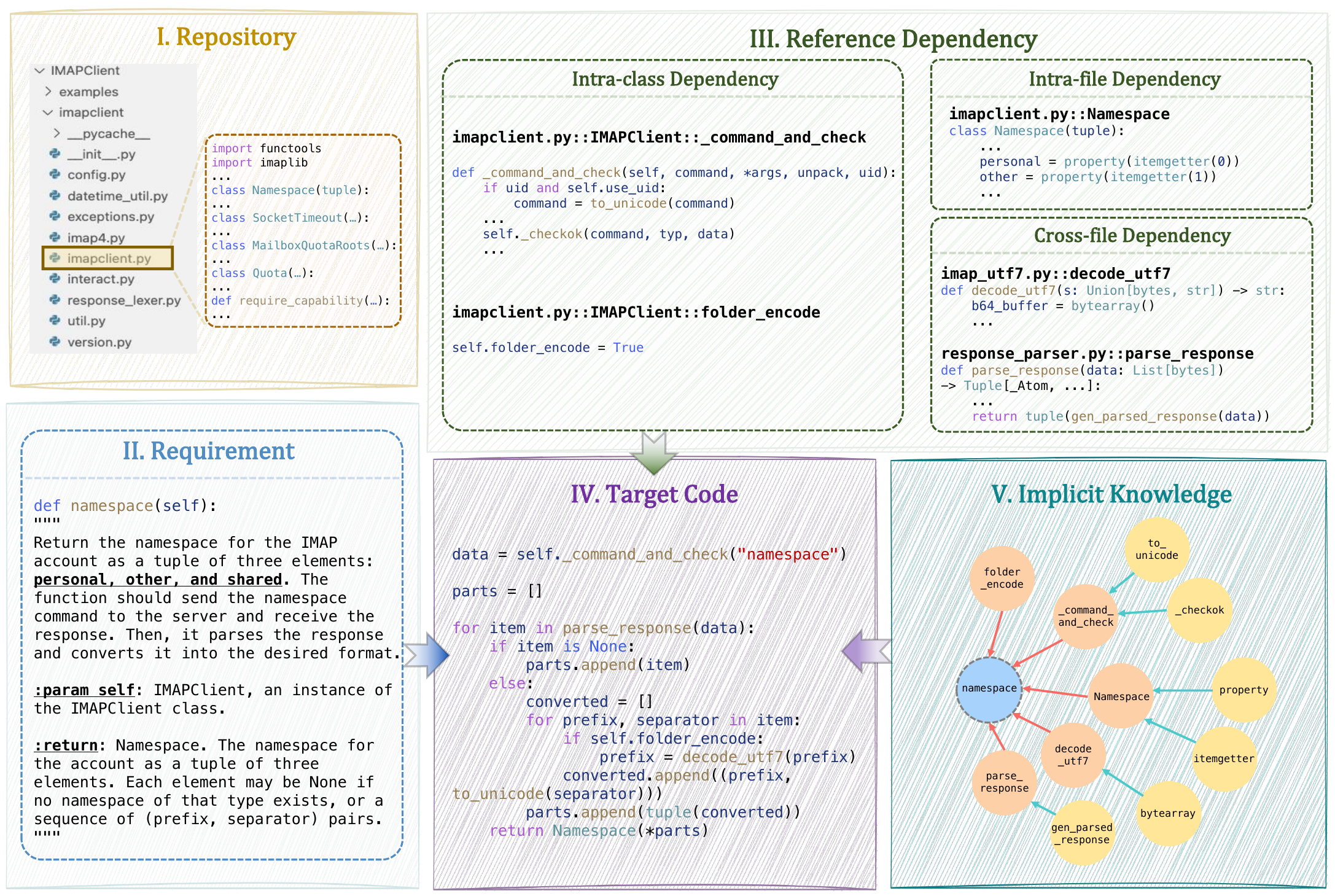}
\caption{An illustration of the repo-level code generation task, as well as the relevant implicit knowledge (V) of the target requirement in the current repository.}
\label{fig: repo-level code generation}
\end{figure*}

To address these challenges, we propose \method, a dual graph-guided LLM agent for retrieval-augmented repo-level code generation. Our approach can bridge the gap between natural language requirements and programming implementations, effectively retrieving implicit code elements, such as invoked APIs and multi-hop related code snippets, and explicit code segments including semantically similar code and up-to-date domain knowledge through web search, even if they are not explicitly expressed in requirements. 
Central to \method are two structured representations: a Requirement Graph (RG) that models requirement relations of code snippets within the repository, as well as the relations between the target requirement and the requirements of these code snippets, and a Structural-Semantic Code Graph (SSCG) that captures both syntactic and semantic relationships within the repository. 
These two graphs are interlinked through mapping relations, enabling node alignment between requirements and code elements. 
Given a natural language requirement, \method first identifies its relevant requirements, such as subrequirements and semantically similar requirements, using the RG. 
These subrequirements are then mapped to corresponding code segments in the SSCG, typically predefined APIs or critical modules that implement core functionalities.  
Because these APIs often depend on other code elements as shown in Figure \ref{fig: repo-level code generation}, \method then employs an LLM-based agent equipped with a unified set of tools designed for structured code navigation and reasoning. The agent performs multi-hop retrieval, traversing the SSCG to gather all contextually relevant code snippets, even those not directly mentioned, based on dependency, invocation, and semantic relations.
In addition to retrieving implicit code dependencies, \method also incorporates semantically similar code snippets from the repository and enhances its context with up-to-date domain knowledge retrieved through targeted web searches.
Figure \ref{fig: reason comparison} shows the retrieved knowledge among \method and other RACG approaches. Most existing RACG approaches fail to retrieval implicit code segments.  
\method can retrieve implicit code including invoked APIs and milti-hop code snippets, and explicit knowledge such as semantically similar code and up-to-date domain information, empowers LLMs to generate accurate, context-aware code that aligns closely with both the stated and latent requirements.

We evaluate \method on three advanced LLMs including non-reasoning models (\ie GPT-4o \cite{GPT-4o} and Gemini-1.5-Pro \cite{team2024gemini}) and reasoning models (\ie QwQ-32B \cite{QwQ}), using the widely-used repo-level code generation benchmarks DevEval \cite{li2024deveval} and CoderEval \cite{yu2024codereval}. Our approach significantly outperforms state-of-the-art (SOTA) baselines across all backbone LLMs. Specifically, \method achieves 43.81\% and 39.15\% relative improvements with GPT-4o and Gemini-1.5-Pro on DevEval, and 31.91\% and 8.25\% relative improvements with GPT-4o and Gemini-1.5-Pro on CoderEval in Pass@1, respectively. Furthermore, on reasoning model QwQ-32B, \method outperforms the best baselines with 10.65\% relative improvements. We also observe that \method can achieve greater improvements on more challenging tasks that require traversing multiple code relationships among different code files, which highlights our approach’s potential for repo-level coding challenges. 
The retrieval process of our approach typically takes only a few seconds on each task, making it highly practical for real-world use.  
The contributions of this paper are as follows:

\begin{itemize}
    \item We propose \method, a dual graph-guided LLM agent for retrieval-augmented repo-level code generation that bridges the gap between natural language requirements and programming implementations, supporting retrieving not only implicit code elements that are not directly mentioned in the requirement but are related to the target code such as invoked APIs and multi-hop code snippets, but also explicit code segments such as semantically similar code and up-to-date domain knowledge with web search.
    
    \item We construct a Requirement Graph (RG) and a Structural-Semantic Code Graph (SSCG) of a repository with mapping relations, meanwhile designing an LLM agent equipped with a unified set of tools, allowing the agent to perform multi-hop retrieval by traversing the SSCG to gather all relevant code snippets, even if they are not directly mentioned in requirements.
    
    \item \method achieves the best performance compared to all baselines at several advanced LLMs on widely-used repo-level code generation benchmarks DevEval and CoderEval. In addition, our approach has more advantage when generating non-standalone target code with complex dependencies, highlighting its potential for repo-level coding challenges.
\end{itemize}

\section{Motivation Example} \label{section: motivation}

In repo-level code generation, given a natural language requirement, the model is asked to output a code snippet to satisfy the requirement. Different from function-level code generation, repo-level code generation is based on a specific repository that contains many predefined code elements such as functions and classes (named APIs) and structural relations. The target code usually invokes these predefined APIs to implement the fine-grained functional elements and seamlessly integrate with the current repository. 
Figure \ref{fig: repo-level code generation} demonstrates an example of the repo-level code generation and shows the relevant code elements of the target requirement in the repository. Given a requirement (II), the model needs to generate the function ``\textit{namespace}'' in the file ``\textit{imapclient.py}''. We can find that the target code (IV) invokes several predefined code elements (III) in the repository (I), including intra-class dependency (\ie ``\textit{$\_$command$\_$and$\_$check}'' and ``\textit{folder$\_$encode}''), intra-file dependency (\ie ``\textit{Namespace}''), and cross-file dependency (\ie ``\textit{decode$\_$utf7}'' and ``\textit{parse$\_$response}''), as presented with orange nodes in V.  
Moreover, successfully generating the target code requires not only
to invoke these code elements above, but also performs multi-step reasoning to understand additional code snippets that interact with or support the implementations of invoked code as shown with yellow nodes in V.

\textbf{However, the natural language requirement only describes a final goal, where these dependencies (\ie orange nodes and yellow nodes as shown in V) are not explicitly mentioned in it.} This disconnect between natural language requirements and programming implementations presents a non-negligible challenge for most existing retrieval approaches \cite{li2023acecoder, li2025large, zhang2024codeagent, liu2024codexgraph}, resulting in the failure to retrieve these code elements. 
To address this challenge, we propose \method, a dual graph-guided LLM agent approach for RACG that supports retrieving implicit code snippets that are not directly mentioned in the requirement but are relevant to the target code such as invoked APIs and multi-hop code snippets, and explicit code elements including semantically similar code and up-to-date knowledge by web search.

\begin{figure}[t] 
\centering
\setlength{\abovecaptionskip}{0.1cm} 
\includegraphics[width=\linewidth]{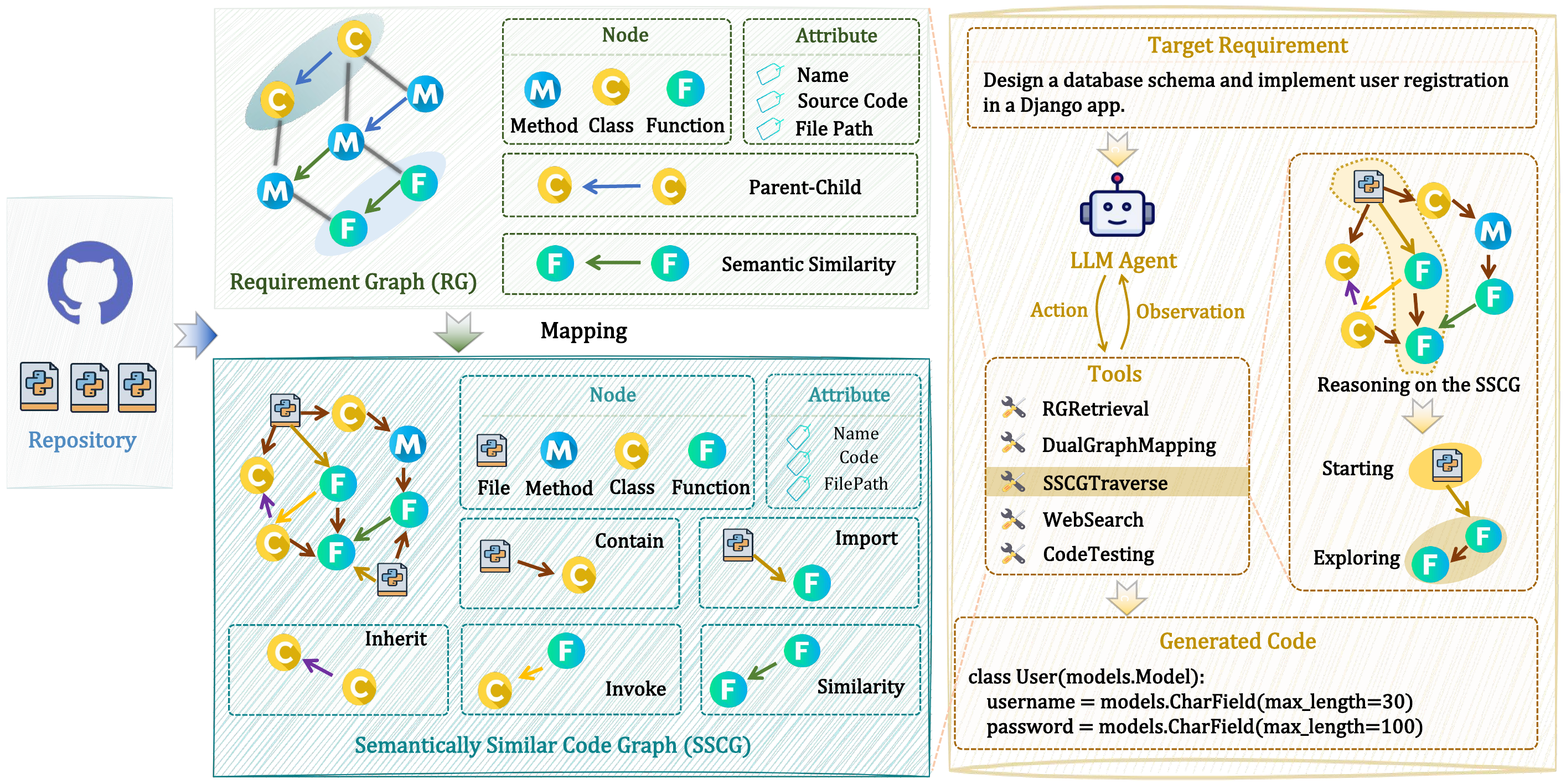}
\caption{Overview of \method. }
\label{fig:workflow}
\end{figure}

\section{\textbfmethod}

\subsection{Overview}

We propose \method, a dual graph-guided LLM-agent approach for retrieval-augmented repo-level code generation, aiming to bridge the gap between natural language requirements and programming implementations. Figure \ref{fig:workflow} shows the overview of \method. Our approach can retrieve not only implicit code snippets that are not directly mentioned in the requirement but are related to the target code such as invoked APIs and multi-hop code elements, but also explicit code elements including semantically similar code and up-to-date external knowledge by web search. 
In the next, Section \ref{sec: re graph} designs a Requirement Graph (RG) that models the requirement relations of code elements within the repository, as well as the relations between
the target requirement and the requirements of these code elements. Section \ref{sec: code graph} introduces a Structural-Semantic Code Graph (SSCG) that captures both syntactic and semantic relationships of code elements within the repository.
Section \ref{sec: agent} presents a dual-graph guided LLM agent through integrating a set of unified tools that enable powerful multi-hop reasoning on SSCG and systematic exploration of the repository to retrieve relevant code contexts.

\subsection{Requirement Graph Construction} \label{sec: re graph}

As described above, a natural language requirement usually describes a final goal, however, it typically comprises multiple fine-grained subrequirements whose code usually has been implemented in the repository and is invoked by the target code, even if these subrequirements are not directly mentioned. 
Besides these subrequirements, the repository typically contains semantically similar elements to the requirement, which is helpful for models to generate the target code.   
Building on this insight, we propose a Requirement Graph (RG) that captures the relations of code elements' requirements in the repository and the relations between the target requirement and the requirements of these code elements, aiming to search for subrequirements and semantically similar requirements of the target requirement.

\subsubsection{RG Definition.}
We model a heterogeneous RG $\mathcal{G_R} (\mathcal{R}, \mathcal{E}, \mathcal{A}, \mathcal{Q})$, where $\mathcal{R} = \left\{ r_i \right\}_{i=1}^{m}$ is node set and $\mathcal{E} \subseteq \mathcal{R} \times \mathcal{R}$ is edge set. Each node $r \in \mathcal{R}$ and edge $e \in \mathcal{E}$ has its type. 
For nodes, $\alpha(r): \mathcal{R}\rightarrow\mathcal{A}$ maps to node types $\mathcal{A} = \left\{\texttt{function requirement}, \texttt{class requirement}, \texttt{method requirement} \right\} $. We set the function level as the smallest node granularity and leave out the file level in node types, which creates a good balance of information density between the index and retrieving requirements. Each node contains several attributes, including its name, source code, and file path. 
For edges, $\beta(e): \mathcal{E}\rightarrow\mathcal{Q} $ maps to edge type $\mathcal{Q} = \{ \texttt{parent-child relation}, \texttt{semantically similar relation} \}$. The $\texttt{parent-child relation}$ means the correlation between a parent requirement and its subrequirement, where the code of the parent requirement usually invokes the code of the child requirement. The $\texttt{semantically similar relation}$ shows that two requirements have similar functionalities.

\subsubsection{RG Construction.}
To construct RG, we first use the static analysis tool tree-sitter\footnote{https://tree-sitter.github.io/} to identify all functions, classes, and methods predefined in the repository. Then, we extract their signatures from predefined code elements and treat functional descriptions within signatures as their requirements if being accessible. For code elements without functional descriptions, we use an advanced LLM (\ie DeepSeek-V2.5 \cite{liu2024deepseek}) with impressive code understanding and generation abilities to generate their requirements, considering that human annotation is time-consuming and labor-intensive. 
After acquiring the descriptions of code snippets, we next extract the relations of requirement nodes. Considering the workload of manual annotation, we also utilize DeepSeek-V2.5 \cite{liu2024deepseek} to annotate
the relationships between requirements. The instruction of acquiring requirements of code snippets and labeling
requirement relations are elaborately designed and shown in the link provided at Section \ref{data link}.
To ensure correctness, we further invite two PhD candidates majoring in computer science to verify and correct all requirements and their relations. 

RG can support to capture related context code that transcends directory boundaries. Two code elements in distant directories (\eg file A and file B) may appear unrelated in traditional navigation, but if they have parent-child relations or share similar semantics, they are syntactically close in our RG. This requirement-level proximity is essential for RACG due to the gap between natural language requirements and programming implementations. By capturing the requirement dependencies, \method effectively identifies related elements even if these elements are not directly mentioned in the requirement.

\subsection{Structural-Semantic Code Graph Construction} 
\label{sec: code graph}

Through RG, our approach can effectively identify the subrequirements and semantically similar requirements of the target requirement, supporting to retrieve the invoked code and functionally similar code of the target code. 
Besides identifying code elements above, it also necessitates reasoning abilities to gather multi-hop contextually relevant code snippets, which interact with or support the implementations of the invoked code, even those not directly mentioned, based on dependency, invocation, and semantic relations. 
To achieve this goal, we build a structural-semantic code graph (SSCG) that captures both syntactic and semantic relations within the repository while maintaining a granularity suitable for balancing retrieved code context length and LLMs' context window length.

\subsubsection{SSCG Definition} Like RG, SSCG is also a heterogeneous directed graph $\mathcal{G_C} (\mathcal{C}, \mathcal{D}, \mathcal{M}, \mathcal{P})$ to index the repository.  $\mathcal{C} = \left\{ c_i \right\}_{i=1}^{n}$ is the node set of code elements and $\mathcal{D} \subseteq \mathcal{C} \times \mathcal{C}$ means the edge set. $\mathcal{M}$ represents the node types $\left\{\texttt{file}, \texttt{class}, \texttt{function}, \texttt{method} \right\} $.
$\mathcal{P}$ means the edge types $\left\{\texttt{import}, \texttt{contain}, \texttt{inherit}, \\ 
\noindent \texttt{invoke}, \texttt{semantical similarity}\right\} $, which represents the relations of two nodes: (1) the $\texttt{import}$ relation from one file to classes or functions in other files; (2) the $\texttt{contain}$ relation from one code element to another, where the one is within another such as class and methods; (3) the $\texttt{inherit}$ relation allows one class to inherit another class; (4) the $\texttt{invoke}$ relation means one code element invokes another code; (5) the $\texttt{semantical similarity}$ relation indicates two code elements have similar semantics. 
For nodes, $\sigma(c): \mathcal{C} \rightarrow \mathcal{M} $ maps each node to its type. For edges, $z(d): \mathcal{D} \rightarrow \mathcal{P} $ maps each edge to its types.

\subsubsection{SSCG Construction}
We extract all programming files of a repository and include them as nodes. Then, we parse each file using the tree-sitter tool to identify predefined functions, classes, and methods in it recursively. Finally, we obtain all predefined code elements of the repository as nodes and acquire the $\texttt{import}$, $\texttt{contain}$, and $\texttt{inherit}$ relations. To obtain the $\texttt{invoke}$ relation, we traverse each code element and use tree-sitter to identify all predefined APIs within it and find these APIs' definitions, assigning the $\texttt{invoke}$ relation between the code element and predefined APIs. Considering that the semantical relation is also essential for agent reasoning to retrieve relevant code contexts besides these structural relations, we incorporate the $\texttt{semantical similarity}$ relation into SSCG. Specifically, we use an advanced embedding model to encode each code element and acquire its vector representation. We then calculate the cosine similarity of two code elements' vector representations and assign an edge if their similarity value is larger than $\epsilon$.
Each node in SSCG has several attributes, including its name, file path, signature, and source code. For storage efficiency, we do not directly store them, rather reserving an index of nodes and edges into Neo4j.

Rather than retrieving only within one file, \method can effectively capture dependency relations among code elements even if they are physically distant in the repository, and enables the agent to navigate the entire repository, supporting retrieving all relevant code contexts through multi-hop reasoning on SSCG. Table \ref{tab:graph comparison} shows the comparison of \method and other graph-based approaches for modeling a repository in various code-related tasks, such as the repository understanding (RepoUnderstander \cite{ma2024understand}), issue localization (OrcaLoca \cite{yu2025orcaloca}), and code localization (LOCAGENT \cite{chen2025locagent}). We can find that our approach starts with the relations of requirements and eliminate the gap between natural language requirements and programs, moreover, comprehensively modeling a repository by introducing diverse node and edge types.

\begin{table*}
  \centering
  \caption{Comparison between \method and other graph-based  approaches for representing a repository in different code-related tasks. Our approach can more comprehensively and effectively models a repository with RG and SSCG that have map relations.  }
    \resizebox{\linewidth}{!}{
    \begin{tabular}{l|cc|ccc|ccccc|cccc}
    \toprule
    \multirow{3}[4]{*}{Approaches} & \multicolumn{5}{c|}{\textbf{Requirement Graph (RG)} }                                       & \multicolumn{9}{c}{ \textbf{Structural-Semantic Code Graph (SSCG)} } \\
   \cmidrule{2-15}        & \multicolumn{2}{c|}{Relation Types} & \multicolumn{3}{c|}{Node Types} & \multicolumn{5}{c|}{Relation Types} & \multicolumn{4}{c}{Node Types} \\
\cmidrule{2-15}        & Parent-Child     & Similarity       &  Class     & Method    & Function  & Contain  & Import  & Inherit    & Invoke    & Similarity  & File & Class & Method  & Function      \\
    \midrule
    CodexGraph \cite{liu2024codexgraph}  &  \xmark     & \xmark  & \xmark & \xmark  & \xmark  &  \cmark & \xmark  & \cmark  &  \xmark   & \xmark   & \xmark &\cmark  &  \cmark  & \cmark  \\
    GraphCoder \cite{liu2024graphcoder} &   \xmark  &  \xmark & \xmark & \xmark  & \xmark  & \xmark & \xmark   & \cmark  & \cmark   & \xmark   &  \xmark  & \cmark & \cmark   & \cmark  \\
    RepoUnderstander \cite{ma2024understand} & \xmark  & \xmark & \xmark  & \xmark   & \xmark  & \cmark    & \xmark  & \cmark   & \cmark   &  \xmark  &  \cmark   &\cmark  & \xmark   & \cmark  \\
   OrcaLoca \cite{yu2025orcaloca}  & \xmark  & \xmark   & \xmark  & \xmark  & \xmark  &  \cmark  &\xmark  & \xmark &  \cmark & \xmark  & \cmark  & \cmark & \xmark  & \cmark \\
  LOCAGENT \cite{chen2025locagent} & \xmark & \xmark  &\xmark  & \xmark  & \xmark  & \cmark  &\cmark &\cmark  & \cmark  &  \xmark  & \cmark   & \cmark   &  \xmark   & \cmark \\
    \midrule
    \method  &  \cmark  & \cmark    &  \cmark &  \cmark & \cmark & \cmark  &  \cmark & \cmark  & \cmark & \cmark  & \cmark  & \cmark & \cmark & \cmark \\
    \bottomrule
    \end{tabular}%
    }
  \label{tab:graph comparison}%
\end{table*}%

\subsection{Dual Graph-Oriented LLM Agent for RACG} \label{sec: agent}

We employ an LLM-based agent equipped with a unified set of tools designed for structured code navigation and reasoning. 
During inference, \method takes the requirement as input and launches the agent that autonomously invokes tools to perform multi-hop retrieval, traversing SSCG to gather all contextually relevant code snippets, even if they are not directly mentioned in the requirement. 
While the agent can iteratively use multiple tools, \method presents a simplified interface to users, which only requires users to submit a requirement and acquire the generated code directly. 
In addition, the RG and SSCG are built offline. These autonomous, easy-to-use ways make our approach both speediness to generate code and effortless to deploy for real-world repo-level coding scenarios.

\begin{table*}
  \centering
  \caption{The input and output of tools designed in \method for repo-level RACG.}
\resizebox{0.9\linewidth}{!}{
    \begin{tabular}{lll}
    \toprule
    Tool Name & \multicolumn{1}{l}{Input} & \multicolumn{1}{l}{Output} \\
    \midrule
    \multirow{2}[1]{*}{\texttt{RGRetrieval} } & \multirow{2}[1]{*}{\textit{Requirement}} &  Subrequirement Nodes $\&$ Semantically \\
          &       &  Similar Requirement Nodes  \\
    \texttt{DualGraphMapping}       &  \textit{Requirement Nodes}   & Starting  Code Nodes  \\
    \texttt{SSCGTraverse}     &  \textit{Starting Code Nodes}    & Traversed Code Nodes   \\
    \texttt{WebSearch}     & \textit{Requirement / Query for Searching}  &  Domain Knowledge \\
    \texttt{CodeTesting}    &  \textit{Generated Code}     &  Testing Results \\
    \bottomrule
    \end{tabular}%
  \label{tab:tools}%
  }
\end{table*}%

\subsubsection{Tool Design for RACG}

Recent studies \cite{zhang2024codeagent, xia2024agentless} have developed a wide array of specialized tools for agents in various software engineering tasks, such as tools with the ability to open, write, and create files. However, these tools are weak in navigating the repository for searching. Building upon our dual graph-based agent approach, we design tools that support efficient repository exploration to retrieve all relevant code contexts. We design five tools in \method described as follows. Table \ref{tab:tools} lists the inputs and outputs of our designed tools in \method.

\texttt{RGRetrieval.} Given a natural language requirement, this tool aims to search its subrequirements and semantically similar requirements from RG. 
Specifically, this tool takes the requirement as input, then uses the RG to find its subrequirement nodes that have \texttt{parent-child relation} with the target requirement node, as well as the semantically similar requirement nodes which are equipped with the \texttt{semantically similar relation} with target requirement node.

\texttt{DualGraphMapping.} This tool constructs the mapping relation between RG and SSCG, supporting mapping the retrieved subrequirement nodes and semantically similar nodes from RG to their corresponding code nodes in SSCG. 
Specifically, we use the attributes of nodes as the mapping index. As described in Section \ref{sec: re graph} and \ref{sec: code graph}, each node has attributes in RG and SSCG, such as its name, file path, and signature. For each retrieved node in RG, this tool concatenates its file path and name as its indexing (\eg \textit{src/imap$\_$utf7.py::decode$\_$utf7}), then uses the indexing to identify its corresponding code node in SSCG. This effectively supports precise mapping from RG to SSCG and prevents the introduction of noise.

\texttt{SSCGTraverse.} This tool performs a navigation on SSCG, initiating from the code nodes acquired from the output of the \texttt{DualGraphMapping} tool and allowing control over both traversal direction and number of hops. This enables the agent to perform multi-hop exploration across the entire repository, retrieving arbitrary code nodes and executing a decision-making action at each step (\ie retrieving or discarding the current node).
Note that by allowing agents to access node types and edge types for each traversal, this tool effectively leverages LLM agents' programming expertise to generate proper meta paths, which is a crucial element for heterogeneous code graph analysis.

\begin{table*}
  \centering
  \caption{Comparison of retrieved knowledge among \method and other RACG approaches, including whether the approach can retrieve implicit knowledge (\ie invoked APIs and multi-hop reasoning code) and explicit knowledge (\ie semantically similar code and up-to-date domain knowledge by web search).}
  \resizebox{\linewidth}{!}{
    \begin{tabular}{lccccc}
    \toprule
    \multirow{2}[2]{*}{\textbf{Approaches}} & \multirow{2}[2]{*}{\textbf{RACG Type}} & \multicolumn{2}{c}{\textbf{Implicit Knowledge}} & \multicolumn{2}{c}{\textbf{Explicit Knowledge}} \\
          &       & \multicolumn{1}{c}{\textbf{Invoked APIs}} & \multicolumn{1}{c}{\textbf{Multi-Hop Code}} & \multicolumn{1}{c}{\textbf{Semantically Similar Code}} & \multicolumn{1}{c}{\textbf{External Knowledge}} \\
    \midrule
    Sparse RACG    & Text-Based RACG    & \xmark   & \xmark   &   \cmark   & \xmark \\
    Dense RACG   & Text-Based RACG     &  \xmark   & \xmark   &  \cmark   & \xmark \\
    RepoCoder \cite{zhang2023repocoder}  & Text-Based RACG     &  \xmark  & \xmark   &  \cmark   & \xmark \\
    GraphCoder  \cite{liu2024graphcoder}   & Graph-based RACG    & \cmark  & \xmark   &   \xmark   & \xmark \\
    CodeAgent  \cite{zhang2024codeagent}   & Agentic RACG     &  \xmark   & \xmark   &  \cmark   & \cmark  \\
    \rowcolor{lightgray} \textbf{\method} & Graph-guided Agentic RACG  & \cmark & \cmark & \cmark & \cmark \\
    \bottomrule
    \end{tabular}%
    }
  \label{tab:knowledge comparison figure}%
\end{table*}%

\texttt{WebSearch.} 
Besides retrieving code snippets from the repository, it is also necessary to retrieve up-to-date domain knowledge through web search. We introduce a web search tool by using a popular search engine DuckDuckGo\footnote{https://duckduckgo.com/}, where programmers often share solutions for programming problems. Compared to other search engines such as Google\footnote{https://www.google.com/} and Bing\footnote{https://www.bing.com/}, DuckDuckGo provides a cheaper and more convenient API.
When encountering a programming question, \method just submits a query to this tool and can acquire useful programming suggestions. Considering that retrieved suggestions are usually long, we then use an LLM to summarize the returned website content as the final tool output. In the process, we block websites that may lead to data leakage. This ensures that \method can retrieve comprehensive domain knowledge for solving the requirement, not being limited to the repository.

\texttt{CodeTesting.} This tool test the generated code, aiming to enhance its correctness and readability. To be specific, we develop Black\footnote{https://github.com/psf/black} as the code test tool, which can check errors of code. 
Given a generated code, this tool checks errors of code such as indentation misalignment and missing keywords.
Subsequently, it tries to rectify these errors and returns the formatted version.

\subsubsection{ReAct Agent Reasoning} 

We apply the ReAct reasoning strategy \cite{yao2023react} to guild the agent in completing retrieval-augmented repo-level code generation. The strategy allows \method to iteratively and autonomously perform actions and observations. Given a requirement, the agent will conduct multiple turns, where each turn predicts an action, invokes the proper tools, treats the output of tools as additional knowledge, and decides whether to generate a final code or invoke other tools for further turns. To be specific, \method systematically performs the following process: (1) For a requirement, \method begins to find out its subrequirement nodes and semantically similar requirement nodes by invoking the \texttt{RGRetrieval} tool, meanwhile employs the \texttt{WebSearch} tool to search relevant domain knowledge if needed. (2) \method invokes the \texttt{DualGraphMapping} tool to map the selected requirement nodes from RG to their corresponding code nodes in SSCG. (3) The agent then invokes the \texttt{SSCGTraverse} tool to navigate SSRG by starting with code nodes acquired in step (2). \method iteratively invokes the knowledge retrieval tools (\ie the \texttt{SSCGTraverse} tool and the \texttt{WebSearch} tool) until it decides to generate code. (4) After predicting the code, \method uses the \texttt{CodeTesting} tool to verify the correctness of generated code. If the code is incorrect, the agent will continually predict actions and invoke tools to correct the code. This process stops until the code is correct or achieves the maximum number of tool invocation. 
Considering that code segments in the local file might be relevant to the target code, we also retrieve the local file content.
Table \ref{tab:knowledge comparison figure} shows the comparison of retrieved knowledge among \method and other RACG approaches. Through the agent reasoning with a unified set of tools, our approach can not only retrieve implicit knowledge such as invoked APIs and multi-hop code, and explicit knowledge including semantically similar code and up-to-date external knowledge.

\section{Evaluation}

\subsection{Research Questions}

We aim to answer the following research questions (RQ): 

\begin{itemize}[leftmargin=*]
    \item \textbf{RQ1: Overall Performance}. How does \method perform on the repo-level code generation task? We conduct experiments on two widely-used repo-level code generation benchmarks (\ie DevEval \cite{li2024deveval} and CoderEval \cite{yu2024codereval}) to evaluate the performance of \method.
    \item \textbf{RQ2: Ablation Study}. What is the contribution of each component in \method? \method retrieves relevant code snippets by integrating a unified set of tools into an agent. In this RQ, we verify the effectiveness of each tool for repo-level code generation. 
    \item \textbf{RQ3: Performance on Different Dependency Types}. 
    What is the performance of \method on different dependency scenarios? Considering that the relevant code contexts of the target code usually come from different directories, this RQ conducts a detailed analysis of \method's ability in different dependency types.
    \item \textbf{RQ4: Generalization Capability}. What does \method perform on reasoning LLMs? Besides evaluating on non-reasoning LLMs, this RQ aims to evaluate our approach's generalization ability on reasoning LLMs.
\end{itemize}

\subsection{Baselines}

We compare \method to six competitive baselines, including text-based RACG, graph-based RACG, and agent-based RACG approaches.

\begin{itemize}[leftmargin=*]
    \item  \textbf{ScratchCG} generates code solely based on the natural language requirement, without incorporating any retrieved code from the repository and other domain knowledge.
    \item  \textbf{Sparse RACG} retrieves textually similar code snippets from the repository. Then the requirement and the retrieved code elements are fed into LLMs. In this paper, we use the BM25 algorithm \cite{robertson1995okapi} to calculate the textual similarity. The smallest-level code snippets are predefined functions, methods, and classes in the repository.
    \item  \textbf{Dense RACG} searches semantically similar code elements from the repository. It uses an embedding model to encode the requirement and code snippets, respectively. Then, it calculates their cosine similarities and retrieves the top-k code elements. 
    \item  \textbf{RepoCoder} \cite{zhang2023repocoder} introduces an iterative retrieval generation pipeline. In each turn, it uses the generated code to retrieve semantically similar code snippets, and then uses the retrieved code as domain knowledge to help models generate code again.  
    \item  \textbf{GraphCoder} \cite{liu2024graphcoder} is a graph-based retrieval-augmented code completion framework that captures the context of completion target through a code context graph (CCG), where CCG consists of control-flow, data-dependency, and control-dependence between code statements. 
    \item  \textbf{CodeAgent} \cite{zhang2024codeagent} is a pioneer LLM-based agent framework for repo-level code generation. It integrates five programming tools, such as the documentation reading tool and symbol navigation tool, and implements four agent strategies to optimize the tools' usage. 
\end{itemize}

To keep a fair comparison, we ensure that the number of retrieved code elements is equal among all baselines and our \method.

\subsection{Dataset}

We evaluate the effectiveness of \method on two widely-used repo-level code generation benchmarks, including DevEval \cite{li2024deveval} and CoderEval \cite{yu2024codereval}. 

\begin{itemize}[leftmargin=*]
    \item  \textbf{DevEval} \cite{li2024deveval} contains 1,825 test examples from 117 repositories, which is significantly larger than the sample sizes compared to many repo-level code generation benchmarks such as CodeAgentBench \cite{zhang2024codeagent}. In addition, DevEval covers popular programming problems and has 10 domains such as Internet and Database. Each test example contains requirements, original repositories, reference code, and reference dependencies. For each example, DevEval requires LLMs to generate code based on the requirement and the current repository. 
    \item  \textbf{CoderEval} \cite{yu2024codereval} originates from open-source projects from various domains. We use the Python tasks of CoderEval to evaluate the effectiveness of \method. It involves functions with contextual dependency and builds a project-level execution platform to provide a ready runtime environment that can automatically assess the functional correctness of generated code. Each example also includes the human-labeled docstring for the target function. 
\end{itemize}

\subsection{Evaluation Metrics}

Following previous works \cite{austin2021program, li2023large, chen2021evaluating, yu2024codereval}, we use Pass@k, a popular metric in code generation as described in Format \ref{format: metrics}. It is the expectation of passing all tests of a task at least once within $k$ attempts, where $n$ is the number of solutions of a task sampled from an LLM and $c$ is the number of correct solutions. In this paper, we use Pass@1 since in real-world scenarios, developers usually only consider the single generated code.

\begin{equation}
\label{format: metrics}
\text{Pass@k} = {{E}}\left[1 - \frac{\binom{n - c}{k}}{\binom{n}{k}}\right]
\end{equation}

\subsection{Implementation Details}

To keep a fair comparison, all experiments requiring embedding use \texttt{stella\_en\_400M\_v5}\footnote{https://huggingface.co/NovaSearch/stella\_en\_400M\_v5} as the embedding model, including identifying semantically similar node relations in RG and SCCG, retrieving semantically similar in Dense RACG and RepoCoder baselines. It is a widely-used and capable embedding model, where embeddings are a numerical representation of a sequence that can be used to measure the relatedness between two sequences. 
Base LLMs used in our experiments, including GPT-4o (GPT-4o-2024-08-06) \cite{GPT-4o}, Gemini-1.5-Pro (Gemini-1.5-Pro-latest) \cite{team2024gemini}, and QWQ-32B \cite{QwQ}, are accessed through their official APIs.  
For each programming task, \method only generates one answer, where the temperature parameter is set to 0. 
To construct the RG and SSCG, we use tree-sitter to parse a repository and acquire all predefined functions, methods, and classes, which are treated as the minimum retrieval units. 
In the RG construction process, we set $\epsilon$ to 0.8. 
In the RACG scenario, the prompt length is a nonnegligible element with the limitation of LLMs' context window. Therefore, if the prompt length is larger than the LLMs' context window length, we truncate the retrieved code snippets from the front and set the maximum generation length of tokens to 500 in baselines, which can support the length of the reference code. The \method and CodeAgent are exceptions, since their multi-turn interaction with tools makes it hard to control the prompt length.
For RepoCoder, we iterate the retrieval generation process two times
since iterating two times achieves the best performance, as demonstrated in its original paper \cite{zhang2023repocoder}.

\begin{table*} 
  \centering
  \caption{The performance of \method and baselines on DevEval and CoderEval. The red color represents the relative improvements of \method compared to the SOTA baseline.}
  \resizebox{0.9\linewidth}{!}{
    \begin{tabular}{l|l|c|cc}
    \toprule
    \multicolumn{1}{l|}{\multirow{2}[2]{*}{Benchmark}} & \multicolumn{1}{l|}{\multirow{2}[2]{*}{Method}} & \multicolumn{1}{c|}{\multirow{2}[2]{*}{RACG Type}} & \multicolumn{2}{c}{Base LLMs} \\
    &  &   & \multicolumn{1}{c}{GPT-4o} & \multicolumn{1}{c}{Gemini-1.5-Pro} \\
    \midrule
    \multicolumn{1}{l|}{\multirow{7}[2]{*}{DevEval \cite{li2024deveval}}} 
    & ScratchCG & \textit{ No Retrieval}   & 17.24   & 14.95 \\
          &   Sparse RACG    & \textit{Text-based } & 27.07  & 36.60 \\
          & Dense RACG & \textit{Text-based } & 40.43 & 39.34 \\
          &    RepoCoder \cite{zhang2023repocoder}  &  \textit{Text-based }  & 30.95   & 30.36 \\
          &   GraphCoder \cite{liu2024graphcoder}     &   \textit{Graph-based }    & 36.44 & 35.67 \\
          &   CodeAgent  \cite{zhang2024codeagent}  &  \textit{Agent-based }      & 28.66  & 33.09 \\
          \rowcolor{cyan!10}
          &   \method    & \textit{Graph-Guided Agent} & \textbf{58.14} (\increase{$\uparrow$ 43.81\%}) & \textbf{54.74} (\increase{$\uparrow$ 39.15\%}) \\
    \midrule 
    \multicolumn{1}{r|}{\multirow{7}[2]{*}{CoderEval \cite{yu2024codereval}}} 
    & ScratchCG & \textit{ No Retrieval}   & 26.96   & 23.91 \\
          &   Sparse RACG    & \textit{Text-based } & 40.87  & 42.17 \\
          & Dense RACG & \textit{Text-based } & 38.26 & 40.43 \\
          &    RepoCoder \cite{zhang2023repocoder}  &  \textit{Text-based }  & 39.13   & 41.30 \\
          &   GraphCoder \cite{liu2024graphcoder}     &   \textit{Graph-based }    & 29.13 & 32.17 \\
          &   CodeAgent  \cite{zhang2024codeagent}  &  \textit{Agent-based }      & 30.00  & 30.43 \\
          \rowcolor{cyan!10}
          &   \method    & \textit{Graph-Guided Agent } & \textbf{53.91} (\increase{$\uparrow$ 31.91\%}) & \textbf{45.65} (\increase{$\uparrow$ 8.25\%}) \\
    \bottomrule
    \end{tabular}%
  \label{tab:main}
  }
\end{table*}%

\section{Results and Analysis}

\subsection{RQ1. Overall Performance }

We conduct experiments on two widely-used repo-level code generation benchmarks (\ie DevEval \cite{li2024deveval} and CoderEval \cite{yu2024codereval}) to evaluate the performance of \method. Table \ref{tab:main} shows the performance of baselines and \method.

\textit{Performance across different benchmarks.}  \method demonstrates an impressive performance advantage on both benchmarks, establishing its superior efficacy for repo-level code generation. On the DevEval benchmark, \method achieves a Pass@1 score of 58.14\% with GPT-4o, marking a 43.81\% relative improvement over the strongest baseline Dense RACG. Similarly, with Gemini-1.5-Pro, it secures the top score of 54.74\%, a relative gain of 39.15 points over the best-performing Dense RACG baseline. This trend of dominance continues on the CoderEval benchmark, where \method obtains Pass@1 scores of 53.91\% (GPT-4o) and 45.65\% (Gemini-1.5-Pro), significantly outperforming all baselines with relative improvements as high as 31.91\% and 8.25\%. These results unequivocally prove that our approach provides the agent with comprehensive and structured code context, effectively guiding them to generate accurate solutions for complex repo-level coding challenges.

\textit{Comparison across different RACG.} A comparative analysis of different RACG paradigms further reveals the source of \method's superiority. Text-based retrieval methods, such as Sparse RACG and Dense RACG, are fundamentally limited as they remain blind to the structural and relational intricacies of a repository, relying solely on lexical or semantic similarity. Meanwhile, the underperformance of CodeAgent suggests that an agent without graph guidance easily gets lost within a complex repository, failing to efficiently explore and locate critical code entities. Furthermore, GraphCoder shows that passively providing a code graph as static context is also insufficient, as the model struggles to actively reason over the graph to find a solution path. In contrast, \method creates a powerful synergy by using the RG and SSCG to bridge the gap between natural language requirements and programming implementations. It supports navigation to steer the agent's active exploration, ensuring that the retrieval process is both goal-driven and contextually aware, thereby overcoming the limitations inherent in other approaches. 

\textit{Performance across diverse base LLMs.} The experimental results underscore the robustness and generalizability of \method across different base LLMs. We observe that the performance of baselines can be inconsistent across LLMs. For instance, the relative ranking of Dense RACG and RepoCoder shifts between GPT-4o and Gemini-1.5-Pro. In stark contrast, \method consistently ranks first. This stability confirms that our approach is a universally effective framework for enhancing the repo-level coding capabilities of LLMs. The fact that the performance gains are often more pronounced with GPT-4o also suggests that more capable models might be better able to reason on SSCG and retrieve the high-quality, structured context.

\vspace{2mm}
\begin{custommdframed}
\textbf{\textit{Answer to RQ1:}} \method outperforms all baselines across all base LLMs on both DevEval and CoderEval, achieving the relative improvements of up to 43.81\% on Pass@1. These results prove the superiority of our dual graph-guided LLM agent through bridging the gap between natural language requirements and programming implementations.
\end{custommdframed}
\vspace{1mm}

\begin{table*}
  \centering
  \caption{Ablation study of \method. We use the complete \method as our baseline and then individually remove (w/o) each tool. ``Usage Time'' means the average usage frequency of each tool during code generation processes. The red color represents the relative decline of each component in \method.}
    \resizebox{0.5\linewidth}{!}{
    \begin{tabular}{lcc}
    \toprule
    Method & \multicolumn{1}{c}{Usage Time} & \multicolumn{1}{c}{GPT-4o} \\
    \midrule
    \rowcolor{cyan!10}
    \method     &  --  & 58.14  \\
    \texttt{w/o  RGRetrieval}    &  1.0   & --  \\
    \texttt{w/o DualGraphMapping}   & 1.0  & --  \\
    \texttt{w/o SSCGTraverse}  &  2.3  & 51.83 (\increase{$\downarrow$ 12.17\%})  \\
   \texttt{w/o WebSearch} &  0.4 & 57.85 (\increase{$\downarrow$ 0.51\%})  \\
    \texttt{w/o CodeTesting}  & 0.8  & 57.09  (\increase{$\downarrow$ 1.84\%})  \\
    \bottomrule
    \end{tabular}%
  \label{tab:ablation}
  }
\end{table*}%

\subsection{RQ2. Ablation Study}

To systematically evaluate the contribution of each key component within \method, we conduct a series of ablation studies. We use the complete \method as our baseline and then individually remove (\texttt{w/o}) each tool. 
All experiments are based on GPT-4o with Pass@1 serving as the performance metric. 
The results are presented in Table \ref{tab:ablation}. 
We also meticulously track the usage frequency of each tool during code generation processes in \method, with the statistics under the column (Usage Time). 
Note that \texttt{RGRetrieval} and \texttt{DualGraphMapping} are the first and second steps of our approach, aiming to identify subrequirement nodes and semantically similar requirement nodes of the target requirement in RG, and then map these nodes into SSCG. 
Only after that, \method can iteratively invoke the \texttt{SSCGTraverse}, \texttt{WebSearch}, and \texttt{CodeTesting} tools. Therefore, in this RQ, we always retain the \texttt{RGRetrieval} and \texttt{DualGraphMapping} tools and only census the usage times of them.

From Table \ref{tab:ablation}, we can observe that the indispensability of graph reasoning is the most significant finding. Removing \texttt{SSCGTraverse} tool, which empowers \method to perform on-demand reasoning and traversal on SSCG, leads to a catastrophic performance relative drop of 12.17\% in Pass@1 (from 58.14 to 51.83). This dramatic decline confirms that the tool facilitates a paradigm shift from static context retrieval to dynamic, on-demand exploration, allowing the agent to navigate complex, cross-file dependencies during code generation.
On average, \method utilizes this tool average 2.3 times per code generation task, a frequency higher than the usage time of other tools. 
The removal of \texttt{CodeTesting} results in a 1.84\% relative decrease (from 58.14 to 57.09), revealing the importance of the self-correction mechanism. This tool functions as a lightweight feedback loop, catching syntactic, or formatting flaws before submission, thereby ensuring code quality. 
Similarly, ablating \texttt{WebSearch} caused a 0.51\% relative drop (from 58.14 to 57.85), highlighting its value in incorporating the external up-to-date knowledge, such as third-party API documentation or common programming patterns. While used less frequently (0.4), it proves decisive in cases requiring information beyond the repository. 

The ablation results confirm that the tools within \method contribute positively to the overall improvement. 
This validates the synergistic design of \method, which designs a unified set of tools designed for bridging the gap between natural language requirements and programming implementations, collectively forming a robust and comprehensive solution for repo-level RACG.

\vspace{2mm}
\begin{custommdframed}
\textbf{\textit{Answer to RQ2:}} Each key component in \method provide positive contributions to the overall improvement, which verifies the synergistic design of our approach. Among these components, the \texttt{SSCGTraverse} tool leads to a catastrophic  relative drop of 12.17\% in Pass@1 (from 58.14 to 51.83) since it shifts from static context retrieval to dynamic, on-demand exploration, allowing the agent to navigate complex, cross-file dependencies.
\end{custommdframed}
\vspace{1mm}

\begin{table*}
  \centering
  \caption{The performance of \method on examples with different dependency types including standalone and non-standalone scenarios.}
    \resizebox{0.95\linewidth}{!}{
    \begin{tabular}{l|c|cccc}
    \toprule
    \multirow{2}[1]{*}{Method} & \multicolumn{1}{r|}{\multirow{2}[1]{*}{Standalone (502)}}  & \multicolumn{4}{c}{Non-standalone (1323)} \\
  \cmidrule{3-6}   &       & Local-file (455) &  Cross-file (157) & Local\&Cross-file (571) & Average \\
    \midrule
    ScratchCG  & \cellcolor{yellow!30}29.28 & 12.08  & 18.47   & 7.88   & \cellcolor{green!20} 9.74 \\
    Sparse RACG     &  \cellcolor{yellow!30}38.44  & 25.93 &  20.38 & 18.39   & \cellcolor{green!20} 21.73  \\
    Dense RACG     & \cellcolor{yellow!30}50.19  & 46.81 & 21.66 &  25.04 & \cellcolor{green!20} 39.79  \\
    RepoCoder \cite{zhang2023repocoder}    & \cellcolor{yellow!30}43.82  & 31.42  &  22.29  & 18.21  & \cellcolor{green!20} 24.07\\
    GraphCoder \cite{liu2024graphcoder}   & \cellcolor{yellow!30}56.97  & 48.79   &  21.58  &32.07  & \cellcolor{green!20} 39.73 \\
    CodeAgent  \cite{zhang2024codeagent}   & \cellcolor{yellow!30}40.63  &31.64  & 18.04  & 18.47  & \cellcolor{green!20}  24.07  \\
    \method    & \cellcolor{yellow!30} 60.16 (\increase{$\uparrow$ 19.86\%})  &  69.67 (\increase{$\uparrow$ 42.79\%}) & 43.31  (\increase{$\uparrow$ 94.30\%}) & 45.18 (\increase{$\uparrow$ 40.88\%})    & \cellcolor{green!20}  48.24 (\increase{$\uparrow$ 21.23\%})  \\
    \bottomrule
    \end{tabular}%
 \label{tab:dependency}
  }
\end{table*}%

\subsection{RQ3. Performance on Different Dependency Types }

To comprehensively evaluate the code generation capability of \method in real-world development scenarios, we further investigate its performance on tasks involving various code dependency relationships. 
Code in practice often has complex and diverse dependencies on other code elements within the repository. Therefore, we categorized the tasks of DevEval into two types based on the dependency relationship between the target code and other code in the repository: Standalone and Non-standalone. 
Standalone (502 examples) refers to examples where the target code's implementation does not depend on any other predefined code elements within the repository. 
Non-standalone (1323 examples) includes cases where the target code needs to invoke other code from the repository, which we further subdivided into three scenarios: Local-file (455 examples), where dependencies are only in the same file with the target code; Cross-file (157 examples), where dependencies are only in other files within the repository; and Local\&Cross-file (571 examples), where dependencies exist both in the local file and across other files. 
Table \ref{tab:dependency} presents Pass@1 results of \method and baselines on different dependency types. The column ``average'' means the performance of different approaches on all non-standalone examples.

The advantage of \method is particularly significant when handling non-standalone code with complex dependencies. While \method achieves the best performance across both standalone and non-standalone categories, the magnitude of its improvement varies dramatically. 
In the standalone category, \method achieves 60.16 Pass@1, a 19.86\% relative improvement over the SOTA baseline (\ie GraphCoder). This suggests that even without code dependencies of the target code, the retrieval mechanism of \method can source valuable domain knowledge, such as specific coding paradigms or data structure definitions, to help LLMs generate higher-quality and project-compliant code. 
In stark contrast, for the non-standalone category, \method achieves an average Pass@1 of 48.24, with a solid 21.23\% relative improvement compared to the best baseline RepoCoder. This substantial lead strongly validates the effectiveness of \method in solving real-world programming tasks, especially when confronted with complex repositories.

For the three types in the non-standalone category, \method achieves an overwhelming victory. 
Specifically, our approach achieves over 40\% relative improvements in all three non-standalone scenarios. 
For the Cross-file category, \method achieves 43.31 Pass@1, nearly doubling the performance of the best baseline RepoCoder with 22.29 Pass@1 (a 94.30\% relative gain). 
We can find that even for the most challenging Local\&Cross-file scenario, \method can acquire the significant improvements, surpassing the best baseline GraphCoder by 40.88\% relative improvements.
We attribute this to the fact that this mixed-dependency scenario best leverages our approach's strength in fusing multi-source information via our dual graph-guided agent, performing multi-hop retrieval, traversing the SSCG to gather all contextually relevant code snippets even if they are in different directories.

\vspace{2mm}
\begin{custommdframed}
\textbf{\textit{Answer to RQ3:}} The advantage of \method is particularly significant when handling non-standalone examples with complex dependencies. Especially, \method acquires the impressive improvements in the most challenging Local\&Cross-file scenario, surpassing the best baseline by 40.88\% relative improvements, which shows that our approach can effectively merge multi-source information via dual graph-guided agent even if they are in various directories.
\end{custommdframed}
\vspace{0mm}

\begin{table*}
  \centering
  \caption{The performance of \method and other RACG baselines on the reasoning model QwQ-32B at the DevEval benchmark.}
    \resizebox{0.58\linewidth}{!}{
    \begin{tabular}{l|c|c}
    \toprule
    Method  & RACG Type  &  QWQ-32B \\
    \midrule
    ScratchCG & \textit{No Retrieval}   & 18.57 \\
    Sparse RACG & \textit{Text-based}  & 34.46 \\
    Dense RACG & \textit{Text-based}  & 47.83 \\
    RepoCoder \cite{zhang2023repocoder} & \textit{Text-based}  & 48.93 \\
    GraphCoder \cite{liu2024graphcoder} & \textit{Graph-based}  & 35.56 \\
    CodeAgent \cite{zhang2024codeagent} & \textit{Agent-based}  & 18.79 \\
        \rowcolor{cyan!10}
    \method  & \textit{Graph-Guided Agent}   & \textbf{54.14} (\increase{$\uparrow$ 10.65\%}) \\
    \bottomrule
    \end{tabular}%
  \label{tab:reasoning}
  }
\end{table*}%

\subsection{RQ4. Generalization Capability }

Recently, reasoning models \cite{QwQ, guo2025deepseek} have shown strong abilities in various code tasks, which have impressive reasoning abilities in the generation process. In this RQ, we explore whether our approach can be applied to reasoning models. Specifically, we evaluate \method on the reasoning model QWQ-32B \cite{QwQ} at DevEval. The results are shown in Table \ref{tab:reasoning}. 

The most striking result is that \method achieves 54.14 Pass@1 on QwQ-32B, achieving the best performance among all RACG approaches. Compared to the directly generation approach (\ie ScratchCG) with 18.57 Pass@1, this represents a substantial absolute improvement of 35.57 points. 
Meanwhile, \method outperforms the SOTA baseline (\ie RepoCoder) with 10.65\% relative improvements.
This remarkable performance leap provides strong evidence that the code generation potential of powerful reasoning models can be significantly unlocked when they are supplied with high-quality, structured contextual information. 
The success of \method on QwQ-32B might be attributed to the following two key factors. First, the advantage of reasoning models lies in their capacity to think step by step. Second, more critically, reasoning models can effectively understand and integrate complex retrieved information to guide the generation process.

\vspace{2mm}
\begin{custommdframed}
\textbf{\textit{Answer to RQ4:}} Besides the non-reasoning models, \method also achieves the best performance compared to the SOTA baseline on reasoning LLM, which demonstrates the satisfying generalization capability of \method.
\end{custommdframed}
\vspace{0mm}

\section{Discussion}

\begin{figure*}[t] 
\centering
\includegraphics[width=\linewidth]{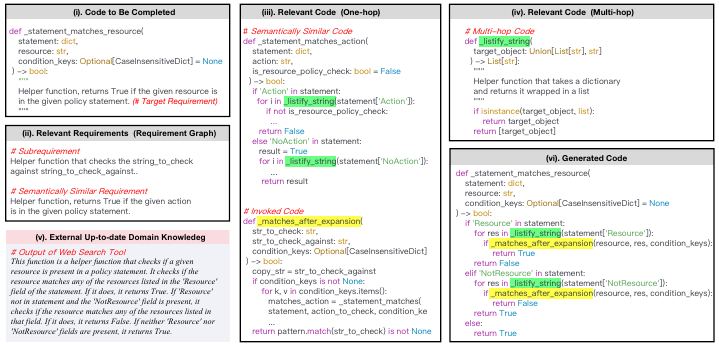}
\caption{An example of the retrieved knowledge and generated code by \method.}
\label{tab: motivation}
\end{figure*}

\subsection{Case Study}

To intuitively illustrate the effectiveness of \method, we present a whole generation workflow of our approach in 
Figure \ref{tab: motivation}, on a practical code generation task. 
This case clearly reveals how \method generates high-quality, repo-consistent code by retrieving relevant code, understanding inter-code dependencies, and finally synthesizing the implementation.

\textbf{(i) Given a target requirement}, the process begins with  \textbf{(ii) identifying its subrequirements and semantically similar ones with the \texttt{RGRetrieval} tool}. Specifically, \method identifies a subrequirement (\textit{Helper function that checks the string\_to\_check against string\_to\_check\_against}) and a semantically similar element (\textit{Returns True if the given action is in the given policy statement}). 
This strategy is superior to monolithic retrieval approaches as it can retrieve relevant requirements even if they are not directly mentioned in the target requirement.
After identifying relevant requirements, \textbf{(iii) \textbfmethod uses the \texttt{DualGraphMapping} tool to map the retrieved requirement nodes in RG into the corresponding code nodes in SSCG}. As shown in Figure \ref{tab: motivation}, the subrequirement is mapped to the function \textit{\_matches\_after\_expansion}, meanwhile the semantically similar requirement is mapped to \textit{\_statement\_matches\_action}. 
Through this way, \method not only finds reusable logic but also establishes a starting point for its subsequent, deeper reasoning process on the SSCG. 
Then, \textbf{ (iv) \textbfmethod begins to perform reasoning on the SSCG to explore contextually relevant code elements through the \texttt{SSCGTraverse} tool}. As shown in Figure \ref{tab: motivation}, our approach retrieves a multi-hop function (\textit{\_listify\_string}) that is invoked by the one-hop semantically similar code (\textit{\_statement\_matches\_action}). This makes sense because when learning the logic and implementation of the semantically similar code to the target code, LLMs need to understand the invoked code elements of the semantically similar code predefined in the repository. 
By traversing the code graph, \method builds a comprehensive context that is not just a collection of isolated snippets but a local code subgraph, encompassing core context code and critical dependencies for generating correct code. It should not be overlooked that \textbf{(v) during the reasoning process, LLMs also use the \textit{WebSearch} tool to search for relevant explanations}. 

Ultimately, \textbf{(vi) \textbfmethod integrates all the retrieved information to generate the target code \textit{\_statement\_matches\_resource}}, including the subrequirement's corresponding code (\textit{\_matches\_after\_expansion}), the semantically similar code (\textit{\_statement\_matches\_action}), the dependent one-hop function (\textit{\_listify\_string}), and supplementary knowledge from web searches (as mentioned in the text). 
The generated code demonstrates a correct logical structure for handling both \textit{Resource} and \textit{NotResource} cases and effectively reuses the retrieved function \textit{\_matches\_after\_expansion} for its core logic, ensuring consistency and reducing redundancy. We can find that based on the retrieved context, \method is able to generate code that is not only functionally correct but also seamlessly integrated and stylistically consistent with the existing repository.

\subsection{Threats to Validity}

\noindent \textbf{Threats to internal validity} include the influence of the model hyper-parameter settings for both our model and the reproduced baselines. To ensure a fair comparison, the number of retrieved code snippets of \method and all baselines keep the same. 
All experiments requiring embedding requirements or code use stella\_en\_400M\_v58 as the embedding model, including identifying semantically similar node relations in RG and SCCG, retrieving semantically similar in Dense RACG and RepoCoder baselines. 
Considering that RepoCoder is designed for retrieval-based code completion, we keep the same setting as its paper and adapt it to the repo-level code generation scenario. We iterate the retrieval generation process two times as recommended in its original paper \cite{zhang2023repocoder}. For all baselines, we parse a repository to acquire all predefined functions, methods, and classes which are treated as the minimum retrieval units. In the code generation process, we set the hyper-parameter (\ie maximum generation length) to 500, which can support the length of the reference codes. Meanwhile, we only generate a single program for each target requirement, where the temperature parameter is set to 0, since developers usually only consider the single generated code in real-world programming environments. Therefore, there is a minor threat to the hyper-parameter settings.

\noindent \textbf{Threats to external validity} include the quality of the benchmarks, baselines, metrics, and base LLMs. We use the widely-used DevEval \cite{li2024deveval} and CoderEval \cite{yu2024codereval} benchmarks to evaluate the effectiveness of \method. 
DevEval is a representative repo-level code generation benchmark that aligns with real-world repositories in multiple dimensions, including code distributions and dependency distributions. It contains 1,825 testing samples from 117 repositories and covers 10 popular domains. Each example is annotated by developers and contains comprehensive annotations (\eg requirements, original repositories, reference code, and reference dependencies).
CoderEval is collected from open-source projects from various domains. Each example involves functions with contextual dependency and builds a project-level execution platform to provide a ready runtime.
To verify the superiority of \method, we consider six advanced RACG approaches, including text-based RACG, graph-based RACG, and agent-based RACG. In addition, to effectively evaluate our approach, we select mainstream LLMs (\ie GPT-4o \cite{GPT-4o} and Gemini-1.5-Pro \cite{team2024gemini}) as base LLMs, and further analyze the performance of \method on reasoning models (\eg QwQ-32B \cite{QwQ}). We apply our approach and baselines to these models and evaluate their performance in repo-level code generation. 
For the metric, following existing studies \cite{li2023acecoder, zhang2024codeagent}, we select a widely used Pass@k metric to evaluate all approaches. It is an execution-based metric that utilizes test cases to check the correctness of generated programs. To ensure fairness, we execute each method two times and report the average experimental results. 
To ensure the reliability of the requirement graph, \method uses the advanced DeepSeek-V2.5 \cite{guo2024deepseek}, an widely-used LLMs with impressive understanding and generation abilities on both natural language and program language, to generate requirements of code snippets and model the relationships between requirements considering being time-consuming and laborious of human labeling. 
In addition, we elaborately design instructions to LLMs, and invite two PhD candidates to manually verify and correct the generated requirements and their relations. We observe that the generated requirements can correctly describe the function of code and effectively predict the relations of requirements. In the future, we will explore more accurate ways to model the requirement graph, even using human annotation methods.

\section{Related Work}

\subsection{Real-world Repo-level Code Generation}

Recently, automated code generation has received continuous attention from researchers \cite{yin2018tranx, wang2025reuse, liu2023syntax, husein2025large}. With the rapid development of LLMs such as DeepSeek \cite{guo2024deepseek}, QwenCoder \cite{qwencoder}, GPT-4o \cite{GPT-4o}, a large body of research \cite{yan2025llm, shrivastava2023repository, tang2024biocoder} has begun to utilize LLMs to generate code for automatic software development, achieving excellent results. 

In the early, code generation tasks mainly focus on generating standalone code units, including statement-level code generation \cite{yin2018learning, austin2021program} and function-level generation \cite{chen2021evaluating, hendrycks2021measuring}. Statement-level code generation requires models to generate a single code statement. Function-level code generation aims to generate a standalone function that can only invoke APIs in the third-party library. 
Despite achieving success, these generated programs are usually short and are independent of other codes. 
For example, each target code in the widely-used benchmark HumanEval \cite{chen2021evaluating} only contains 11.5 lines and 24.4 tokens on average. 
In software development, human programmers typically work within a code environment. They extend their functionalities based on the foundational code framework such as a code repository.

Inspired by this, some studies \cite{zhang2024codeagent, li2024deveval} introduce the repo-level code generation task. Given a code repository, the repo-level code generation task aims to generate code based on the software artifacts included in the repository, encompassing the requirements, code dependencies, and runtime environment. 
This growing interest has led to the emergence of several benchmarks, such as RepoEval \cite{zhang2023repocoder}, CoderEval \cite{yu2024codereval}, DevEval \cite{li2024deveval}, EvoCodeBench \cite{li2024evocodebench}, and SWEBench \cite{jimenez2023swe}, further advancing the field.
Different from simple code generation tasks,  repo-level code snippets usually have intricate contextual dependencies, which are too complex for existing LLMs to handle and generate.
Meanwhile, researchers \cite{zhang2024codeagent, liu2024codexgraph} also explore to improve the performance of repo-level code generation through designing an iterative retrieval generation pipeline \cite{zhang2023repocoder}, constructing the graph for the repository \cite{liu2024codexgraph}, and introducing an agentic framework allowing LLMs to use external programming tools \cite{zhang2024codeagent}. 
In this paper, we propose a dual graph-guided LLM agent for enhancing the performance of models in real-world repo-level code generation.

\subsection{Retrieval-Augmented Code Generation}

In recent years, LLMs have demonstrated impressive capabilities in code generation tasks. For instance, where GPT-4o and DeepSeek-V3 can achieve 92.7\% and 91.5\% Pass@1 performance on HumanEval \cite{chen2021evaluating}. However, their performance often degrades significantly when turned to repository-level code generation tasks within complex, large-scale repositories. 
To eliminate this challenge, retrieval-augmented code generation (RACG) approaches have been introduced, aiming to provide LLMs with relevant contextual information.
Given a requirement, RACG retrieves contextually relevant code and integrates the retrieved information into the generation process \cite{ridnik2024code, li2023codeie, fakhoury2024llm, zhang2024codedpo}. 
Existing RACG approaches are mainly divided into three categories, including text-based RACG, graph-based RACG, and agent-based RACG.

Text-based RACG focuses on textual similarity to retrieval, which encompasses sparse retrievers such as using the BM25 algorithm to calculate the textual similarities between the requirement and candidate codes, and embedding-based dense retrievers \cite{li2023acecoder, su2024dragin} which invoke embedding models to encode them and select semantically similar code. Different from these approaches, RepoCoder \cite{zhang2023repocoder}  introduces an iterative retrieval generation pipeline. In each turn, it uses the generated code to retrieve semantically similar code snippets from the repository, and then uses the retrieved code as domain knowledge to help models generate code again.
Though achieving improvements, text-based RACG overlooks the structured nature of programming languages, where repositories cover intricate dependencies that is valuable information for retrieval.
To eliminate this challenge, some researchers \cite{liu2024graphcoder, liang2024repofuse, liu2024evaluating, liu2024codexgraph} have introduced graph-based RACG by modeling the dependency relations of code. Among them, GraphCoder \cite{liu2024graphcoder} constructs the code context graph (CCG) \cite{wei2024selfcodealign} that considers the control-flow, data-flow, and control-dependence between code statements for retrieval-augmented code completion.
CODEXGRAPH \cite{liu2024codexgraph} utilizes static analysis to extract code graphs from repositories. In the graph, nodes represent source code symbols such as module, class, and function, and edges between nodes represent the relationships among these symbols. However, this approach is limited by graph query syntax and still can not bridge the gap between natural language and programming implementations.
Recently, agent-based approaches have received increasing attention in code generation due to their powerful tool utilization capability. 
Agent-based RACG integrates retrieval into the reasoning process, allowing LLMs to interact with repositories \cite{zhang2024codeagent, xia2024agentless, chen2025locagent}. CodeAgent \cite{zhang2024codeagent} is a pioneer LLM-based agent framework for repo-level code generation. It allows LLMs to invoke external programming tools such as web searching, and designs four agent strategies to optimize these tools’ usage. Subsequently, researchers introduce several agentic frameworks for other real-world coding tasks. For example, Agentless \cite{xia2024agentless} is designed for solving issues by preprocessing the repository’s structure and file skeleton, allowing LLMs to interact with contextually relevant code. LocAgent \cite{chen2025locagent} addresses code localization through combining the graph-based repository representation and the agentic machine, enabling LLM agents to effectively search and locate relevant entities.

Different from existing approaches, \method designs a dual graph-guided LLM agent RACG framework that bridges the gap between natural language requirements and programming implementations. It supports retrieving implicit code snippets including invoked APIs and multi-hop code, and
explicit code snippets such as semantically similar code and up-to-date domain knowledge, even if they are not directly mentioned in the requirement.

\section{Conclusion}

In this paper, we propose \method, a dual graph-guided LLM
agent for RACG, bridging the gap between natural language requirements and programming implementations.
Central to \method are two structured representations: a Requirement Graph (RG) and a Structural-Semantic
Code Graph (SSCG) to model the structure of the repository.
Based on this,  \method effectively retrieves implicit code elements, such as invoked APIs and multi-hop related code snippets, and explicit code segments including semantically similar code and up-to-date domain knowledge through web search, even if they are not explicitly expressed in requirements. 
Extensive experiment results demonstrate \method significantly outperforms the state-of-the-art baselines on repo-level code generation. Moreover, \method can achieve greater improvements on more challenging tasks that require traversing multiple code relationships among different code files, highlighting its potential in real-world coding challenges.

\section{Data Availability} \label{data link}
The code implementation and data are publicly available at figshare anonymous link\footnote{https://figshare.com/s/4148a1c56d08804cd75a}.

\bibliographystyle{ACM-Reference-Format}
\bibliography{main}

\end{document}